\documentclass{llncs}

\usepackage{psfig}

\long\def\comment#1{}
\newcommand{\compact}{\vspace{-0.6ex}}

\usepackage{psfig,pifont,xspace}
\usepackage{citesort}
\usepackage{amssymb}

\usepackage{mathrsfs}

\newcommand{\IR}{{\rm\hbox{I\kern-.15em R}}}
\newcommand{\IN}{{\rm\hbox{I\kern-.15em N}}}
\newcommand{\IZ}{{\sf\hbox{Z\kern-.40em Z}}}

\newenvironment{algotext}[2]
      {\begin{minipage}{1.0\textwidth}
       \raggedright\noindent
       {\bf Input:} #1\\
       {\bf Output:} #2\\
       \rule[3pt]{1.0\textwidth}{0.3pt}%
       \setlength{\leftmargini}{1.5em}%
       \setlength{\leftmarginii}{2.5em}%
       \setlength{\leftmarginiii}{1em}%
       \vspace{-3.0ex}\begin{description}}
      {\end{description}\end{minipage}}

\newcommand{\step}{\vspace{-0.0ex}\item}

\begin{document}

\title{Highly Scalable Algorithms for Robust String Barcoding\thanks{
The work of BD was supported in part by NSF grants CCR-0206795, CCR-0208749
and NSF CAREER grant IIS-0346973.
The work of IIM was supported in part by a ``Large Grant'' from the University 
of Connecticut's Research Foundation.}}

\author{Bhaskar DasGupta\inst{1}
\and Kishori M. Konwar\inst{2}
\and Ion I. M\u{a}ndoiu\inst{2}
\and Alex A. Shvartsman\inst{2}
}

\institute{Department of Computer Science, 
University of Illinois at Chicago\\ 
Chicago, IL 60607-7053, \email{dasgupta@cs.uic.edu}
\and
Computer Science and Engineering Department, 
University of Connecticut\\
371 Fairfield Rd., Storrs, CT 06269-2155, 
\email{kishori,ion,aas@cse.uconn.edu}
}

\date{}

\maketitle

\begin{abstract}
String barcoding is a recently introduced 
technique for genomic-based identification of microorganisms.
In this paper we describe the engineering of highly scalable 
algorithms for robust string barcoding. 
Our methods enable distinguisher selection based on 
{\em whole genomic sequences} of hundreds of microorganisms of up 
to bacterial size on a well-equipped workstation, 
and can be easily parallelized to further 
extend the applicability range to thousands of bacterial size genomes. 
Experimental results on both randomly generated and NCBI 
genomic data show that whole-genome based selection results in 
a number of distinguishers nearly matching the information theoretic 
lower bounds for the problem.
\end{abstract}

\section{Introduction}
\label{intro.sec}

String barcoding is a recently introduced 
technique for genomic-based identification of 
microorganisms such as viruses or bacteria. 
The basic barcoding problem \cite{RG02} is formulated as follows:
given the genomic sequences $g_1,\ldots g_n$ of $n$ microorganisms, 
find a minimum number of strings $t_1,\ldots,t_k$ {\em distinguishing}
these genomic sequences, i.e., having the property that,
for every $g_i\neq g_j$, there exists a string $t_l$ which
is a substring of $g_i$ or $g_j$, but not of both. A 
closely related formulation was independently proposed in
\cite{BCVFJ01}, where it is assumed that it is possible to detect not just the 
presence or absence of a distinguisher $t_i$, 
but also the number of repetitions of $t_i$ as a substring, up to 
a threshold of $R>0$. The formulation in \cite{RG02}, which we adopt 
in this paper, corresponds to $R=1$.

Identification is performed by spotting or synthesizing 
on a microarray the Watson-Crick complements of the 
distinguisher strings $t_1,\ldots,t_k$, 
and then hybridizing to the array the fluorescently labeled DNA 
extracted from the unknown microorganism. 
Under the assumption of perfect hybridization stringency, 
the hybridization pattern can be viewed as a 
string of $k$ zeros and ones, referred to as the {\em barcode} 
of the microorganism.
By construction, the barcodes corresponding to the $n$ microorganisms  
are distinct, and thus the barcode uniquely identifies
any one of them.
To improve identification robustness, one may also require 
{\em redundant distinguishability} 
(i.e., at least $m$ different distinguishers for every pair of microorganisms, 
where $m>1$ is some fixed constant) 
and impose a lower bound on the edit distance 
between any pair of selected distinguishers \cite{RG02}.

The algorithms previously proposed for string barcoding 
are based on integer programming \cite{RG02}, and on 
Lagrangian relaxation and simulated annealing \cite{BCVFJ01}. 
Unfortunately, the run-time of these algorithms does not scale well with 
the number of microorganisms and the length of the genomic sequences, 
e.g., the largest instance sizes reported in \cite{RG02} 
have a total genomic sequence length of around 100,000 bases. 

In this paper we describe the engineering of highly scalable 
algorithms for robust string barcoding. 
Our methods enable distinguisher selection based on 
{\em whole genomic sequences} of hundreds of microorganisms of up 
to bacterial size on a well-equipped workstation, 
and can be easily parallelized to further 
extend the applicability range to thousands of bacterial size genomes. 
Whole-genome based selection is beneficial in at least two significant ways.
First, it simplifies assay design since the DNA of the unknown 
pathogen can be amplified using inexpensive general-purpose 
whole-genome amplification methods such as
specialized forms of degenerate primer multiplex PCR \cite{CheungN96}
or multiple displacement amplification \cite{Dean02}. 
Second, whole-genome based selection results in a reduced number of 
distinguishers, often very close to the information theoretic 
lower bound of $\lceil\log_2{n}\rceil$.

Our algorithms are based on a simple greedy selection strategy -- 
in every iteration we pick a substring that distinguishes the 
largest number of not-yet-distinguished pairs of genomic 
sequences.  This selection strategy
is an embodiment of the greedy setcover algorithm (see, e.g., \cite{Vazirani})
for a problem instance with $O(n^2)$ elements corresponding to 
the pairs of sequences.  
Hence, by a classical result of \cite{Chv79,Joh74,Lov75}, 
our algorithm guarantees an approximation factor of 
$2\ln n$ for the barcoding problem.
Very recently, Berman et al. \cite{BDK03} have shown that 
no approximation algorithm can guarantee a factor of 
$(1-\epsilon)\ln{n}$ unless $NP=DTIME(n^{\log\log{n}})$, 
and also proposed an information content greedy heuristic 
achieving an approximation factor of $1+\ln n$.
Experimental results given in Section \ref{results.sec} show 
that our setcover greedy algorithm produces solutions of virtually 
identical quality to those obtained by the information content 
heuristic.  

The setcover greedy algorithm is extremely versatile, and 
can be easily extended 
to handle redundancy and minimum edit distance constraints, 
as well as other biochemical constraints on individual distinguisher sequences. 
Furthermore, unlike the information content heuristic of \cite{BDK03}, 
the greedy setcover algorithm can also take into account 
genomic sequence uncertainties expressed in the form of degenerate bases.  
Although degenerate bases are ubiquitous in genomic databases, 
previous works have not recognized the need to properly handle them.
For example, 
experiments in \cite{RG02} have implicitly treated degenerate bases 
in the input genomic sequences as distinct nucleotides; under 
this approach a substring of degenerate nucleotides such as 
NNNNN, might be erroneously selected as a distinguisher although 
it encodes for any possible substring of length 5.

To achieve high scalability, our implementation relies on 
several techniques. First, we use an incremental algorithm for quickly 
generating a representative set of candidate distinguishers 
and collecting all their occurrences in the given genomic sequences.  
To reduce the number of candidates, we avoid generating 
any substring that appears in all genomic sequences, which 
typically eliminates very short candidates. For each genomic 
sequence, we also generate only one of the substrings 
that appear exclusively in that sequence, this optimization 
eliminates from consideration most candidate distinguishers 
above a certain length.  
Unlike the suffix tree method proposed by Rash and Gusfield \cite{RG02}, 
our approach may generate multiple candidates 
that appear in the same set of $k$ genomic sequences (for $1<k<n$). 
However, the penalty of having to evaluate redundant 
candidates in the candidate selection phase is offset in practice by the 
faster candidate generation time. 
Finally, the efficient implementation of the greedy selection phase of algorithm 
combines a partition based method for computing the 
coverage gain of candidate distinguishers (this method was first 
proposed in the context of the information content heuristic in 
\cite{BDK03})
with a ``lazy'' strategy for updating coverage gains.

The rest of the paper is organized as follows.
In Section \ref{formulations.sec} we give formal problem formulations 
and review previous work.  
In Section \ref{greedy_implementation.sec} we describe 
the efficient implementation of the setcover greedy algorithm 
for the basic string barcoding problem.
In Section \ref{extensions.sec} we discuss the modifications 
required in the implementation for handling degenerate bases 
in input genomic sequences, redundancy and edit distance 
constraints, as well as biochemical constraints such as 
constraints on melting temperature and GC-content.
In Section \ref{results.sec} we give the results of a comprehensive 
experimental study comparing, on both randomly generated 
and genomic data, our setcover greedy algorithms with 
other scalable methods including the information content heuristic 
and a recent set multicover randomized rounding approximation 
algorithm.  

\comment{
We conclude in Section \ref{conclusions.sec} with directions 
for further research.}


\section{Preliminaries and Problem Formulation}
\label{formulations.sec}

Let $\Sigma=\{a,c,g,t\}$ be the DNA alphabet, and 
$\Sigma^*$ be the set of string over $\Sigma$. 
A {\em degenerate base} is a non-empty subset of $\Sigma$.  
We identify degenerate bases of cardinality 1 with the 
respective non-degenerate bases. Given a DNA string 
$x=x_1\ldots x_k\in \Sigma^*$ and 
a string of degenerate bases $y=y_1\ldots y_n$, $n\ge k$, 
we say that 
\begin{itemize}
\compact
\item $x$ has a {\em perfect match} at position $i$ of $y$ iff 
$y_{i+j-1} = \{x_j\}$ for every $1\le j\le k$, 
\item $x$ has a {\em perfect mismatch} at position $i$ of $y$ iff 
there exists $1\le j\le k$ such that $\{x_j\}\not\subseteq y_{i+j-1}$, and  
\item $x$ has an {\em uncertain match} at position $i$ of $y$ iff 
$\{x_j\}\subseteq y_{i+j-1}$ for every $1\le j\le k$, but 
$y_{i+j-1}\neq \{x_j\}$ for at least one $j$.
\compact
\end{itemize}
String $x=x_1\ldots x_k\in \Sigma^*$ 
{\em distinguishes} two sequences of degenerate bases $y$ and $z$ 
iff (a) $x$ has a perfect match at one or more positions of $y$, and 
has perfect mismatches at all positions of $z$, or, symmetrically, 
(b) $x$ has a perfect match at one or more positions of $z$, and 
has perfect mismatches at all positions of $y$.
\comment{
\smallskip 

\noindent
{\bf String barcoding problem}\\
{\bf Given:} sequences of degenerate bases $g_1,\ldots g_n$\\
{\bf Find:} minimum number of strings $t_1,\ldots,t_k\in \Sigma^*$ such that, 
for every $i\neq j$, there exists a string $t_l$ distinguishing 
$g_i$ and $g_j$.

\smallskip 
}
The robust string barcoding problem with degenerate bases is formulated as follows:
{\sl Given sequences of degenerate bases $g_1,\ldots g_n$ and redundancy threshold $m$, 
find a minimum number of strings $t_1,\ldots,t_k\in \Sigma^*$ such that, 
for every $i\neq j$, there exist $m$ distinct strings $t_l$ distinguishing 
$g_i$ and $g_j$.}

It is easy to see that, for $m=1$, at least $\lceil\log_2{n}\rceil$ distinguishers are 
needed to distinguish any $n$ genomic sequences. However, achieving this lower bound 
requires distinguishers that have perfect matches in nearly half of the 
sequences. In practice, additional constraints, such as lower bounds on 
the length of distinguishers, may result in no string having perfect matches 
in a large number of sequences, and therefore much more than a logarithmic number 
of distinguishers.
The next theorem, the proof of which we omit due to space constraints, 
establishes under a simple probabilistic model 
that there is an abundance of distinguishers perfectly matching 
at least a constant fraction of the input sequences.

\begin{theorem}\label{result4}
Consider a randomly generated instance of the string barcoding problem 
over a fixed alphabet $\Sigma$ in which there are $n$ strings, each string 
$s=s_0s_1\ldots s_{\ell-1}$ is
of length exactly $\ell$ generated independently randomly with 
$Pr[s_i=a]=1/|\Sigma|$ for any $i$ and any $a\in\Sigma$. 
Also assume that $\ell$ is sufficiently large compared to $n$. 
Then, 
for a random string $x\in\Sigma^*$ of length $O(\log\ell)$, 
the expected number of the input strings which contain $x$ as a 
substring is $pn$ for some constant $0<p<1$.
\end{theorem}

\begin{proof}
Assume $n$ and $\ell$ to be sufficiently large for asymptotic results
and $\sigma=|\Sigma|>1$ to be fixed.
It suffices to show that
for a random string $x\in\Sigma^*$ of length $k=O(\log\ell)$, 
$Pr[\mbox{$x$ is a substring of $s$}]=p$ for some constant $0<p<1$ and 
$s$ is any one of the input $n$ strings.
In~\cite[Examples 6.4, 6.7, 6.8, 9.3 and 10.11]{O95}, 
Odlyzko uses the bounds and generating function described 
in~\cite{GO81} to give asymptotic bounds on 
$Pr[\mbox{$x$ is a substring of $s$}]$
when
$\sigma=2$. 
The result can be generalized to the case of
any fixed $\sigma>2$ as follows.
For a fixed $x=x_1x_2\ldots x_k$, define the correlation 
polynomial $C_x(z)$ of $x$ as $C_x(z)=\sum_{j=0}^{k-1}c_x(j)z^j$ 
where $c_x(0)=1$ and, for $1\leq j<k$, 
\compact
\[
c_x(j)=\left\{\begin{array}{ll} 1 & 
\mbox{if $x_1x_2\ldots x_{k-j}=x_{j+1}x_{j+2}\ldots x_k$} \\ 
0 & \mbox{otherwise} \\ \end{array}\right.
\]
\compact
Let $f_x(\ell)$ be the number of strings in $\Sigma^*$ of length $\ell$ 
that do not contain $x$ as a substring and 
$F_x(z)=\sum_{\ell=0}^\infty f_x(\ell)z^\ell$
be the generating function for this number. 
Then, $F_x(z)=\frac{C_x(z)}{z^k+(1-\sigma z)C_x(z)}$. 
From this, it follows that
$
Pr[x\prec s]=1-{\mathbf{e}}^{-\frac{\ell}{\sigma^k C_x(1/\sigma)}
		+O(\ell k \sigma^{-\sigma k})}
             + O({\mathbf{e}}^{-\ell/O(1)})
$
for all sufficiently large $n,k$ and $\ell$, where ${\mathbf{e}}$ is the 
base of natural logarithm.
Note that $1\leq C_x(\sigma)<2$ and for a specific $x$, 
$C_x(\sigma)$ can be calculated exactly. Now,
setting $k=\Theta(\log_\sigma\ell)$ gives 
$Pr[\mbox{$x$ is a substring of $s$}]=p$ 
for some constant $0<p<1$. $\qed$
\end{proof}

\compact
\noindent
{\bf Previous work.}
The robust string barcoding problem was introduced (for the case when 
genomic sequences contain no degenerate bases) by Rash          
and Gusfield~\cite{RG02}; they provided some                
experimental results based on integer programming methods, 
and left open the exact complexity and approximability of this problem.
The problem without redundancy constraints was independently 
considered by Borneman {\em et al.}~\cite{BCVFJ01}, who 
also considered non-binary distinguishability (based on detecting 
the multiplicity of a distinguisher as a substring)
and a slightly more general problem in which the objective is to 
pick a given number of distinguishers maximizing the number of 
distinguished pairs. 
The main motivation for the formulations in \cite{BCVFJ01}
comes from minimizing the number of oligonucleotide
probes needed for analyzing populations of ribosomal RNA gene (rDNA)
clones by hybridization experiments on DNA microarrays.
Borneman {\em et al.} provided computational results using 
Lagrangian relaxation and simulated annealing techniques, 
and noted that the problem is NP-hard assuming that
the lengths of the sequences in the prespecified set were unrestricted.
Very recently, 
Berman, DasGupta and Kao~\cite{BDK03} considered a general
framework for test set problems that captured the string 
barcoding problem and its variations; their main contribution
is to establish theoretically matching lower and upper 
bounds on the worst-case approximation ratio.    
Cazalis et al. \cite{Cazalis04} have independently 
investigated similar greedy distinguisher selection strategies 
for string barcoding.  Unlike our work, 
the algorithms in \cite{Cazalis04} consider only a small 
random subset of the possible distinguishers and also prescribe 
their length in order to achieve practical running time.

\section{Efficient Implementation of the Greedy Setcover Algorithm}
\label{greedy_implementation.sec}

In this section we present the implementation 
of the setcover greedy algorithm in the context of the basic string 
barcoding problem, i.e., we disregard redundancy constraints
and the presence of degenerate bases in the input sequences. 
Implementation modifications needed to handle the 
robust barcoding problem in its full generality are discussed 
in Section \ref{extensions.sec}.

Our implementation of the setcover greedy algorithm has two main phases:
a {\em candidate generation phase} and a {\em candidate selection phase}. 
In the candidate generation phase a representative set of candidate 
distinguishers is generated from the given genomic sequences.
For each generated candidate, we also compute the list of sequences with 
which the candidate has perfect matches; this information is needed in 
the candidate selection phase. 
To reduce the number of candidates, we avoid generating 
any substring that appears in all genomic sequences, which 
typically eliminates very short candidates. For each genomic 
sequence, we also make sure to generate only one of the substrings 
that appear exclusively in that sequence, this optimization 
eliminates from consideration most candidate distinguishers 
above a certain length.  
Unlike the suffix tree method proposed by Rash and Gusfield \cite{RG02}, 
our approach may generate multiple candidates 
that appear in the same set of $k$ genomic sequences (for $1<k<n$). 
However, the penalty of having to evaluate redundant 
candidates in the candidate selection phase is offset in practice by the 
faster candidate generation time.

Efficient implementation of the above candidate elimination rules 
is achieved by generating candidates in increasing order of length  
and using exact match positions for candidates of length $l-1$ when generating 
candidates of length $l$. 
For each position $p$ in the input genomic sequences, we also maintain a 
flag to indicate whether or not the algorithm should evaluate 
candidate substrings starting at $p$.  The possible 
values for the flag are TRUE (the substring of current length starting 
at $p$ is a possible candidate), FALSE (we have already saved 
the substring of current length starting at $p$ as a candidate), 
or DONE (all candidates containing as prefix the substring of 
current length starting at $p$ are redundant, i.e., the position 
can be skipped for all remaining candidate lengths). 
Initially all flags are set to TRUE.  The FALSE flags are reset to 
TRUE whenever we increment the candidate length, however, 
we never reset DONE flags.
For every candidate length $l$, candidate evaluation proceeds 
sequentially over all positions of the genomic sequences. 
Whenever we reach a position $p$ whose flag is set to TRUE, 
we use the list of matches for the substring of length
$l-1$ starting at $p$ (or a linear time string matching algorithm if $l$ 
is the minimum candidate length) to determine the list of matches 
for the substring of length $l$ starting at $p$, and set 
the flag to FALSE for all positions where these matches occur. 
If the substring of length $l$ starting at $p$ 
has matches only within the source sequence, and we have already 
generated a ``unique'' candidate for this sequence, we discard 
the candidate and set the flag of $p$ to DONE.

\begin{figure}[t]
\fbox{\small
\begin{algotext}
{Set $C$ of candidate distinguishers}
{Set $D$ of selected distinguishers}
\step $D \gets \emptyset$; For every $c\in C$, $\Delta_{old}(c)\gets \infty$
\step Repeat 
\begin{description}
\step $\Delta^*\gets 0$
\step For every $c\in C$ with $\Delta_{old}(c) > \Delta^*$ do \quad\quad // Since $\Delta(c,D)\le \Delta_{old}(c)$, $c$ can be ignored if $\Delta_{old}(c)\le \Delta^*$
\begin{description}
\step 
$\Delta_{old}(c) \gets \Delta(c,D)$
\step If $\Delta(c,D) > \Delta^*$ then $\Delta^*\gets \Delta(c,D)$; $c^*\gets c$
\end{description}
\step
If $\Delta^* > 0$ then $D\gets D\cup\{c^*\}$
\end{description}
\step While $\Delta^* > 0$ 
\end{algotext}
}
\caption{
\label{barcode_greedy_algo.fig} The greedy candidate selection algorithm}
\compact
\compact
\compact
\compact
\compact
\end{figure}

A further speed-up technique is to generate candidate
distinguishers from a strict subset of the input sequences.
Although this speed-up can potentially affect solution quality,
the results in Section \ref{results.sec} show that the solution 
quality loss for whole-genome barcoding is minimal, 
even when we generate candidates based on a single input 
sequence, which corresponds to pre-assigning a
barcode of all 1's to this sequence.

After the set of candidates is generated we select the final 
set of distinguishers in the greedy phase of the algorithm 
(Figure \ref{barcode_greedy_algo.fig}). 
We start with an empty set of distinguishers $D$.
While there are pairs of sequences that are not yet distinguished by $D$, 
we loop over all candidates  
and compute for each candidate $c$ the number $\Delta(c,D)$
of pairs of sequences that are distinguished by $c$ but not by $D$, 
then add the candidate $c$ with largest $\Delta$ value to $D$.
Two sequences $s$ and $s'$ are distinguished by 
a candidate $c$ iff exactly one of $s$ and $s'$ appears 
in the list $P_c$ of perfect matches of $c$, which is available 
from the candidate generation phase. 
A simple method for computing $\Delta$ values 
is to maintain an $n\times n$ symmetric matrix indicating which 
of the pairs of sequences are already distinguished, and then 
to probe the $|P_c|\cdot (n-|P_c|)$ entries in this matrix corresponding to 
pairs $(s,s')$ with $s\in P_c$ and $s'\notin P_c$ when computing $\Delta(c,D)$. 
A more efficient method is 
based on maintaining the partition defined on the set of sequences 
by $D$.
If the partition defined by $D$ consists of sets 
$S_1,\ldots,S_k$, then 
we can compute $\Delta(c,D)$ in $O(k+|P_c|)=O(n)$ time using the 
observation that 
\compact
\begin{equation}\label{delta.eq}
  \Delta(c,D) = \sum_{i=1}^k |S_i\cap P_c|\cdot |S_i\setminus P_c|
\end{equation}
\compact
In addition to the fast partition based computation, 
our implementation of the greedy selection phase 
uses a lazy strategy for updating the $\Delta$ values, 
based on the observation that they are monotonically 
non-increasing during the algorithm
(see Figure \ref{barcode_greedy_algo.fig}).

\section{Extended Barcoding Requirements}
\label{extensions.sec}

In this section we describe the modifications needed 
to the basic implementation given in previous section when 
handling practical extensions of the barcoding problem.

\noindent
{\bf Degenerate bases.} 
In the presence of degenerate bases in the 
input genomic sequences, the hybridization of a particular 
distinguisher may depend on which bases are actually present 
positions with degeneracy greater than 1.  
The greedy setcover algorithm takes into account this 
possibility for uncertain hybridization by only counting 
a pair $(g,g')$ as distinguished by a candidate $c$ if and only if 
$c$ has a perfect match with one and only perfect mismatches with the other. 
For each generated candidate, 
in addition to the list of sequences that have only perfect matches 
we also save a list containing all sequences with at least one 
uncertain match.  This allows fast computation of the 
(typically much longer) list of sequences having only perfect 
mismatches.  
To avoid generating candidate distinguishers containing degenerate bases, 
we set the DONE flag as soon as the corresponding substring extends past a 
degenerate base. 
Finally, since the partition of genomic sequences is no longer defined 
in the presence of uncertain hybridization, formula (\ref{delta.eq})
is no longer applicable and we have to use the 
$n\times n$ ``distinguished so far'' matrix for computing $\Delta$ values.

\noindent
{\bf Biochemical constraints on individual distinguishers.}
Since selected distinguishers must hybridize under the same experimental 
conditions, in practice it is natural to impose a variety of constraints 
on individual distinguishers, such as minimum and  maximum length, 
GC content, melting temperature, etc. Furthermore, we may want to 
avoid using as distinguishers strings which appear in other 
organisms that may contaminate the sample.  All individual constraints 
are easily incorporated as a simple filter in the candidate generation 
phase.  

\noindent
{\bf Redundancy constraints and minimum edit distance constraints.} 
In practice, robust identification requires redundant distinguishability, 
i.e., more than one distinguisher distinguishing any given pair of 
genomic sequences.  One may also impose a lower bound on the edit distance
between any pair of selected distinguishers \cite{RG02}.
Taking into account redundancy requirements is done by maintaining 
the number of times each pair of genomic sequences has been distinguished.
In order to incorporate the minimum edit distance constraint, 
after selecting a distinguisher we eliminate from consideration all 
candidates that  are within an edit distance smaller than the given 
threshold.

\section{Experimental Results}
\label{results.sec}

We performed experiments on both randomly generated  
instances and whole microbial genomes extracted from the NCBI databases \cite{ncbi}.
Random testcases were generated from the uniform distribution 
induced by assigning equal probabilities to each of the four 
nucleotide; these testcases do not contain any nucleotides 
with degeneracy greater than 1. 
The NCBI testcase represents a selection of 29 complete 
microbial sequences, varying in length between 490,000 
and 4,750,000 bases (over 76 million bases in total). 
All experiments were run on a PowerEdge 2600 Linux
server with 4 Gb of RAM and dual 2.8 GHz Intel Xeon CPUs -- only one of
which is used by our sequential algorithms.


\subsection{Algorithm Scalability}

\begin{figure}[t]                 
\centerline{\psfig{figure=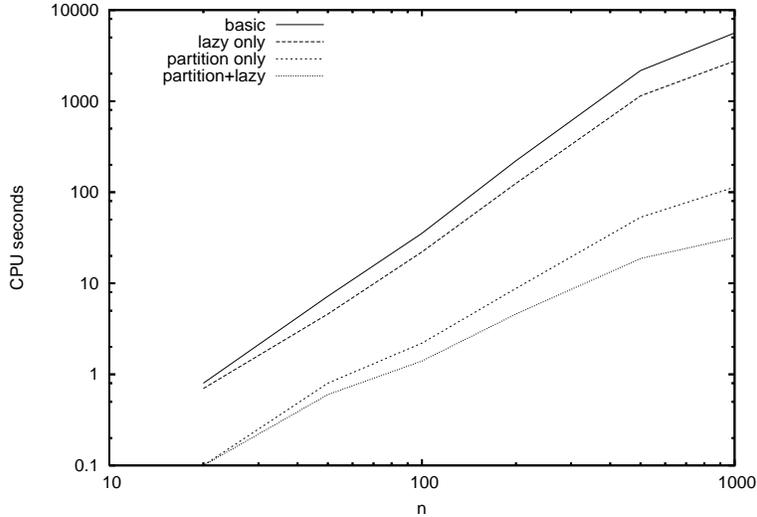,width=4.0in}}
\compact
\caption{
\label{greedy-speed-up.fig}
Candidate selection CPU time (in seconds) for 
$n$ random sequences of length 10,000 and redundancy 1, 
averaged over 10 instances of each size.}
\end{figure}

As described in Section \ref{greedy_implementation.sec}, 
there are two main phases in the algorithm: 
candidate distinguisher generation, and greedy 
candidate selection. 
Figure \ref{greedy-speed-up.fig} gives 
the average candidate selection CPU time for              
$n$ random sequences of length 10,000 and redundancy 1, 
averaged over 10 instances of each size.
Combining the two 
speed-up techniques for this phase (partition based coverage 
gain computation and lazy update of candidate gains) 
results in over two orders of magnitude reductions 
in runtime.

\setlength{\tabcolsep}{3pt}

\begin{table}[t]
\begin{center}
{\footnotesize
\caption{
\label{sources.tab} Average statistics for instances 
with 1,000 random sequences of length 10,000, redundancy 1, and number 
of source sequences varying from 1,000 down to 1.}
\begin{tabular}{|c|rrrrrrrr|}
\hline
\#Source Seq. 
   & 1000 & 50 &10 &5 &4 &3 &2 &1\\
\hline
\#Candidates & 
       7213568.8 &      1438645.4 &     402700.5 &      225842.7 &     186871.3 &       146054.3 &      102800.6 &      55738.9 \\
\#Matches & 
        55696250.9 &     35246584.9 &    23162063.6 &    18384468.1 &   16898179.9 &     15037610.7 &    12532936.0 &    8741587.0 \\
Gen. time & 
        132.3 &  44.7 &  35.5 &  31.4 &  31.3 &  30.6 &  28.1 &  24.9 \\
Selection time &
        31.7 &  10.7 &  5.3 &   3.6 &   3.4 &   3.1 &   2.3 &   1.6 \\
\#Distinguishers &
        14.1 &  14.1 &  14.1 &  14.1 &  14.0 &  14.1 &  14.2 &  14.5 \\
\hline
\end{tabular}
}
\end{center}  
\compact
\compact
\compact
\end{table}

As mentioned in Section \ref{greedy_implementation.sec}, 
a further speed-up technique is to generate candidate 
distinguishers only from a small number of ``source'' input sequences.
Table \ref{sources.tab} gives the average number of candidates, 
number of matches, runtimes for candidate generation and 
greedy selection, and number of selected distinguishers 
for instances with 1,000 random sequences of length 10,000 and redundancy 1, 
when the number of source sequences is varied from 1,000 down to 1 
(the source sequences were chosen at random). 
Although this speed-up can potentially affect solution quality, 
we found that on large instances the solution quality loss is minimal 
even when we generate candidates based on a 
single input sequence; this case corresponds to pre-assigning a 
barcode of all 1's to the source sequence.  
The technique reduces significantly both the memory requirement 
(which is proportional to 
the number of candidates and the number of times they match
input sequences) and the runtime required for 
candidate generation and greedy selection. 
\comment{
As shown in Table \ref{million.tab}, this makes the method applicable 
to hundreds of sequences of bacterial genome size on a 
well-equipped workstation.
}

\setlength{\tabcolsep}{6pt}

\comment{

\begin{table}
\begin{center}
{\footnotesize
\caption{
\label{million.tab} Average statistics for instances 
with up to 100 random sequences of length 1,000,000 and redundancy 1
(number of source sequences set to 1).}
\begin{tabular}{|c|rrrcc|}
\hline
$n$ & \#Candidates & \#Matches & Gen. time & Select time & \#Distinguishers \\
\hline
10    &  2039766.8 &     8281127.2 &     45.1 &  0.6 &   4.0 \\
20    &  2607128.9 &     16730749.0 &    87.0 &  1.3 &   5.0 \\
50    &  3363016.8 &     43802244.9 &    216.1 & 3.6 &   6.6 \\
100   & 3921359.6 &     91346850.3 &    444.4 & 8.5 &   8.0 \\
\hline
\end{tabular}
}
\end{center}  
\end{table}

}


\begin{table}[b]
\compact
\compact
\compact
\begin{center}
{\footnotesize
\caption{\label{quality1} Number of distinguishers returned by the 
setcover greedy algorithm (SGA) for varying redundancy and number of 
sequences.
For each value of $n$ we report the average over 10 testcases, 
each consisting of $n$ random sequences of length 10,000.
For comparison we include information content heuristic 
results (ICH) and the information theoretic lower bound of 
$\lceil\log_2{n}\rceil$ for redundancy 1 (LB). }
\begin{tabular}{|c|r|rrrrrrr|}
\hline
Algorithm & r & 
  $n=10$ & $n=20$ & $n=50$ & $n=100$ & $n=200$ & $n=500$ & $n=1000$ \\
\hline
LB& 
1 & 4 & 5 & 6 &  7 & 8 & 9 & 10 \\
ICH & 
1&  4.0 & 5.0 & 7.0 & 8.0 & 10.0 & 12.2 & 14.1 \\
SGA & 
1 & 4.0 & 5.0 & 7.0 & 8.0 & 10.0 & 12.3 & 14.1 \\
SGA & 
2   &    6.7 &
        8.3 &
        10.6 &
        12.5 &
        14.1 &
        16.7 &
        18.9 \\
SGA & 
3  &     8.8 &
        11.6 &
        13.6 &
        15.5 &
        17.3 &
        20.1 &
        22.4 \\
SGA & 
4   &   10.8 &
        14.0 &
        16.5 &
        18.7 &
        20.7 &
        23.5 &
        26.1 \\
SGA & 
5   &    13.6 &
        16.6 &
        19.5 &
        21.5 &
        23.7 &
        26.8 &
        29.5 \\
SGA & 
10  &   22.5 &
        26.8 &
        32.0 &
        34.6 &
        37.5 &
        41.7 &
        44.9 \\
SGA & 
20  &   43.0 &
        47.6 &
        55.6 &
        59.5 &
        63.4 &
        68.0 &
        72.6 \\
\hline
\end{tabular}
}
\end{center}  
\end{table}

\subsection{Solution Quality on Random Data}

Table \ref{quality1} gives the number of distinguishers 
returned by the setcover greedy algorithm for 
redundancy varying between 1 and 20 on between 10 and 
1,000 random sequences of length 10,000.
For comparison we include in the table the results 
obtained by the information content heuristic 
results of \cite{BDK03}, as well as the information 
theoretic lower bound of $\lceil\log_2{n}\rceil$
for the case when the redundancy requirement is 1.
We note that the number of distinguishers 
returned by the setcover greedy algorithm is 
virtually identical to that returned by the 
information content heuristic, despite the latter 
one having a better approximation guarantee \cite{BDK03}. 
Furthermore, the results for redundancy one 
are within 50\% of the information 
theoretic lower bound for the range of instance sizes 
considered in this experiment.  The gap between 
the solutions returned by the algorithms and the 
lower bound does increase with the number of sequences; 
however it is not clear how much of this increase is caused by 
degrading algorithm solution quality, and how much by 
degrading lower bound quality.

We also compared our setcover greedy algorithm with a 
recently proposed multi-step rounding algorithm for set multicover \cite{BDS04}.  
The rounding algorithm has the following steps:
\begin{enumerate}
\compact
\item Solve the fractional relaxation
of the natural integer program formulation of problem \cite{RG02}
(we used the commercial solver CPLEX 9.0 for implementing this step)
\item Scale the fractional solution by an appropriate constant factor 
(see \cite{BDS04} for details)
\item Deterministically select all distinguishers with a 
scaled fractional value exceeding 1
\item Randomly select a subset of the remaining candidates, 
each candidate being chosen with a probability equal to 
the scaled fractional value
\item Finally, if the selected set of distinguishers is not yet feasible, 
add further distinguishers using the setcover greedy algorithm
\compact
\end{enumerate}
The approximation guarantee established in \cite{BDS04} for the 
general set multicover problem translates into an approximation 
factor of $2\ln{n}-\ln{r}$ for robust string barcoding 
with redundancy $r$, which suggests that 
the multi-step rounding algorithm is likely to 
improve upon the setcover greedy for high redundancy constraints.
Table \ref{quality2} gives the results of experiments 
comparing the setcover greedy and multi-step rounding algorithms 
on testcases consisting of up to 200 random sequences, each
of length 1,000 for redundancy requirement ranging from 1 to 100. 
The results confirm that the multi-step rounding algorithm 
has better solution quality than setcover greedy when 
redundancy requirement is very large relative to the number of sequences
(entries typeset in boldface), yet the setcover greedy still 
has the best performance for most combinations of parameters.

\begin{table}[t]
\begin{center}
{\footnotesize
\caption{
\label{quality2} Number of distinguishers returned by the 
setcover greedy algorithm (SGA) and the 
multi-step rounding algorithm in \cite{BDS04} (RND)
for varying redundancy and number of sequences.
For each value of $n$ we report the average over 10 testcases, 
each consisting of $n$ random sequences of length 1,000.
Boldface entries correspond to instances for which the multi-step 
rounding algorithm has better solution quality than setcover greedy.}
\begin{tabular}{|c|r|rrrrr|}
\hline
Algorithm & r & 
  $n=10$ & $n=20$ & $n=50$ & $n=100$ & $n=200$  \\
\hline
\hline
SGA &1 &        4.0 &   5.0 &   7.0 &   9.0 &   11.0 \\
RND &1 &        5.0 &   6.8 &   10.5 &  13.0 &  16.0 \\
\hline
SGA &2 &        6.3 &   8.2 &   11.2 &  12.9 &  15.0 \\
RND &2 &        7.3 &   10.7 &  14.8 &  17.0 &  20.4 \\
\hline
SGA &5 &        13.2 &  16.1 &  19.5 &  22.4 &  24.6 \\
RND &5 &        13.2 &  18.2 &  23.5 &  27.3 &  31.2 \\
\hline
SGA &10 &       22.8 &  27.0 &  32.1 &  36.1 &  39.4 \\
RND &10 &       {\bf 20.2} &  30.9 &  37.4 &  41.9 &  48.3 \\
\hline
SGA &20 &       43.4 &  48.8 &  57.0 &  61.0 &  65.8 \\
RND &20 &       {\bf 38.9} &  50.7 &  62.6 &  69.4 &  76.2 \\
\hline
SGA &50 &       100.9 & 112.0 & 125.6 & 133.8 & 142.0 \\
RND &50 &       {\bf 92.6} &  {\bf 107.8} & {\bf 125.2} & 141.6 & 159.5 \\
\hline
SGA &100 &      195.0 & 217.2 & 239.0 & 255.5 & 264.0 \\
RND &100 &      {\bf 184.9} & {\bf 205.2} & {\bf 236.0} & 270.0 & 289.0 \\
\hline
\end{tabular}
}
\end{center}  
\compact
\compact
\compact
\end{table}


\subsection{Experiments on Genomic Data}

We ran our algorithm on a set of 29 complete
microbial genomic sequences extracted from NCBI databases 
\cite{ncbi}.
Sequence lengths in the set vary between 490 Kbases and 4.75 Mbases, 
with an average length of 2.6 Mbases (over 76 Mbases total).
Unlike random testcases, the sequences in the NCBI data set contain 
a small number of degenerate bases, 861 bases in total. Therefore, 
we cannot use the partition method for computing the number of 
sequence pairs distinguished by a candidate in the greedy selection phase, 
and we have to use the slower matrix datastructure. 
In these experiments we varied the redundancy requirement from 
1 to 20.  To see the effect of length and edit distance 
requirements on the number of distinguishers, 
for each redundancy requirement we computed 
both an unconstrained solution, and a solution in which 
distinguishers must have length between 15 and 40, 
and there should be a minimum edit distance of 6 between 
every two selected distinguishers (these values are 
similar to those used in \cite{RG02}.
In all experiments, we generated candidates based 
only on the shortest sequence of 490 Kbases.

The results on the NCBI dataset are given in Table \ref{ncbi.tab}.
Naturally, meeting higher redundancy constraints 
requires more distinguishers to be selected. 
Additional length and edit distance constraints 
further increase the number of distinguishers, 
but the latter is still within reasonable limits.
The length constraints reduce the number of 
candidates (from 1,775,471 to 122,478), 
which, for low redundancy values has the effect of reducing 
greedy selection time.  
However, for high redundancy requirements the reduction 
in number of candidates is offset by the increase 
in solution size, and greedy selection becomes more time consuming 
with length and edit distance than without (selection time 
grows roughly linearly with solution size).

\begin{table}[t]
\begin{center}
{\footnotesize
\caption{
\label{ncbi.tab} Results on a set of 29 NCBI complete microbial genomes.
Candidate generation time is approximately 335 seconds for all 
combinations of parameters.}
\begin{tabular}{|c|rrrcc|}
\hline
Redundancy & $l_{min}$ & $l_{max}$ & MinEdit & Select time & \#Distinguishers \\
\hline
1 &0 & $\infty$ &0 & 14.2 &  6.0 \\
1 &15 &40 &6  &  2.6 &   8.0 \\
\hline
5 &0 & $\infty$ &0 & 20.3 &  21.0 \\
5 &15 &40 &6  &  8.7 &   31.0 \\
\hline
10 &0 & $\infty$ &0  &      22.9 &  41.0 \\
10 &15 &40 &6 &  16.4 &  60.0 \\
\hline
20 &0 & $\infty$ &0   &     26.8 &  76.0 \\
20 &15 &40 &6 &  33.4 &  123.0 \\
\hline
\end{tabular}
}
\end{center}  
\compact
\compact
\compact
\end{table}

\section{Conclusions}
\label{conclusions.sec}

In this paper we have given highly scalable 
algorithms for the robust string barcoding problem, 
and have shown that distinguisher selection based 
whole genomic sequences results in a number of 
distinguishers nearly matching 
the information theoretic lower bounds for the problem. 

In ongoing work we are exploring heuristics
and approximation algorithms for several 
extensions of the string barcoding problem. 
First, we are considering the use of probe mixtures as distinguishers. 
With most microarray technologies it is feasible to spot/synthesize 
a mixture of oligonucleotides at any given microarray location. 
The DNA of a pathogen  will hybridize to such a location 
if it contains at least one substring which is the Watson-Crick complement 
of one of the oligonucleotides in the mixture.
Using oligonucleotide mixtures as distinguishers can reduce the 
number of spots on the array -- and therefore barcode length -- closer 
to the information theoretical lower-bound of $\log_2 n$.  
The reduction promises to be particularly significant when reliable 
hybridization requires relatively long distinguishers; 
in these cases even the optimum barcoding length is far from  
$\log_2 n$ \cite{RG02}.
A special case of this approach is the use of 
{\em degenerate} distinguishers similar to the degenerate 
primers that have been recently employed in 
multiplex PCR amplification \cite{LinhartS02,SouvenirBSZ03}.
Degenerate distinguishers are particularly 
attractive for string barcoding since their synthesis cost is nearly 
identical to the synthesis cost of a single non-degenerate 
distinguisher (synthesis requires the same number of steps, 
the only difference is that multiple 
nucleotides must be added in some of the synthesis steps). 

In many practical pathogen identification applications 
collected biological samples may contain the DNA of multiple pathogens.
This issue is considered to be particularly significant 
in medical diagnosis applications,
see, e.g., \cite{GKNAULA03} for studies in
detecting more than one HPV (human papiloma virus) genotype with
varying rate of multiple HPV infections carried by the same HPV carrier. 
In future work we plan to develop extensions of the barcoding 
technique that can reliably detect multiple pathogens 
for a given bound on the number of pathogens present.

\section*{Acknowledgments}
The authors would like to thank Claudia Prajescu for her 
help with the implementation of the multi-step rounding 
algorithm in \cite{BDS04}.

{\footnotesize
\bibliographystyle{plain}
\bibliography{longnames,barcode}
}
\end{document}